\documentclass[useAMS,usenatbib]{mn2e}

\usepackage{epsfig}

\title[MUNICS: Connecting star formation and stellar mass]{The
connection between star formation and stellar mass: Specific star
formation rates to redshift one\footnotemark[2]\footnotemark[3]}

\author[Georg Feulner et al.]
{Georg~Feulner$^1$\footnotemark[1], 
Yuliana~Goranova$^{1,2}$,  
Niv~Drory$^{3}$, 
Ulrich~Hopp$^{1,2}$, \newauthor
Ralf~Bender$^{1,2}$\vspace*{.25em}
\\
$^1$Universit\"ats-Sternwarte M\"unchen, Scheinerstra\ss e 1, D--81679
M\"unchen, Germany\\
$^2$Max-Planck-Institut f\"ur Extraterrestrische Physik,
Giessenbachstra\ss e, D--85748 Garching bei M\"unchen, Germany\\
$^3$University of Texas at Austin, Austin, Texas 78712
}

\begin{document}

\date{Accepted 2004 November 23. Received 2004 November 18; in
  original form 2004 October 6}

\pagerange{\pageref{firstpage}--\pageref{lastpage}} \pubyear{2004}

\maketitle

\label{firstpage}

\begin{abstract}
We investigate the contribution of star formation to the growth of
stellar mass in galaxies over the redshift range $0.5 \, < \, z \, <
\, 1.1$ by studying the redshift evolution of the specific star
formation rate (SSFR), defined as the star formation rate per unit
stellar mass. We use an $I$-band selected sample of 6180 field
galaxies from the Munich Near-Infrared Cluster Survey (MUNICS) with
spectroscopically calibrated photometric redshifts. The SSFR decreases
with stellar mass at all redshifts. The low SSFRs of massive galaxies
indicates that star formation does not significantly change their
stellar mass over this redshift range: The majority of massive
galaxies have assembled the bulk of their mass before redshift
unity. Furthermore, these highest mass galaxies contain the oldest
stellar populations at all redshifts. The line of maximum SSFR runs
parallel to lines of constant star formation rate. With increasing
redshift, the maximum SFR is generally increasing for all stellar
masses, from $\mathrm{SFR} \simeq 5 \: M_\odot \: \mathrm{yr}^{-1}$ at
$z \simeq 0.5$ to $\mathrm{SFR} \simeq 10 \: M_\odot \:
\mathrm{yr}^{-1}$ at $z \simeq 1.1$. We also show that the large SSFRs
of low-mass galaxies cannot be sustained over extended periods of
time. Finally, our results do not require a substantial contribution
of merging to the growth of stellar mass in massive galaxies over the
redshift range probed. We note that highly obscured galaxies which
remain undetected in our sample do not affect these findings for the
bulk of the field galaxy population.
\end{abstract}


\footnotetext[1]{E-mail: feulner@usm.lmu.de}

\footnotetext[2]{Based on observations collected at the Centro
  Astron\'omico Hispano Alem\'an (CAHA), operated by the
  Max-Planck-Institut f\"ur Astronomie, Heidelberg, jointly with the
  Spanish National Commission for Astronomy.}

\footnotetext[3]{Based on observations collected at the VLT (Chile)
  operated by the European Southern Observatory in the course of the
  observing proposals 66.A-0123 and 66.A-0129.}

\section{Introduction}
\label{s:intro}

\begin{figure*}
\epsfig{figure=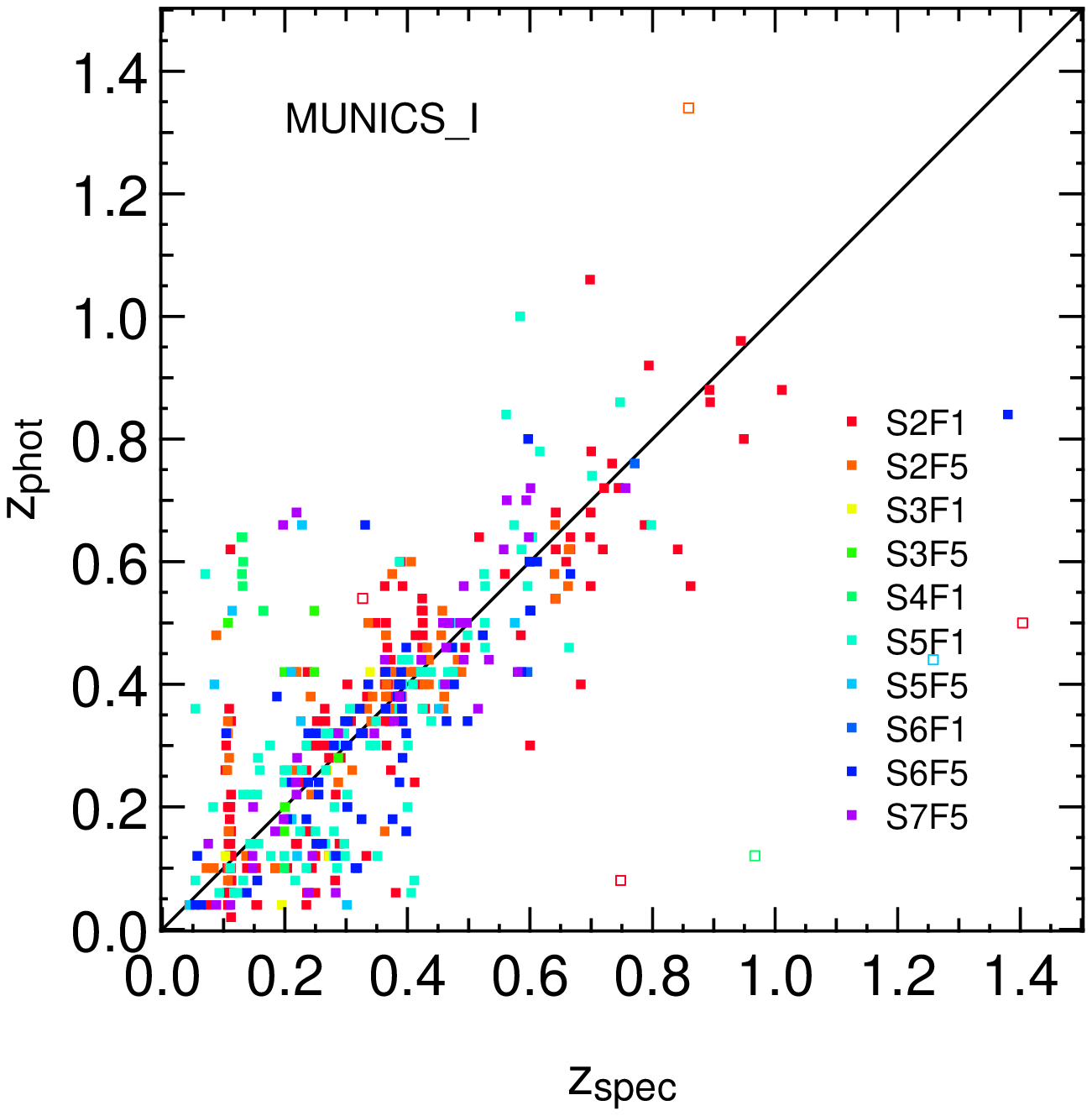,width=0.31\textwidth}
\hfill
\epsfig{figure=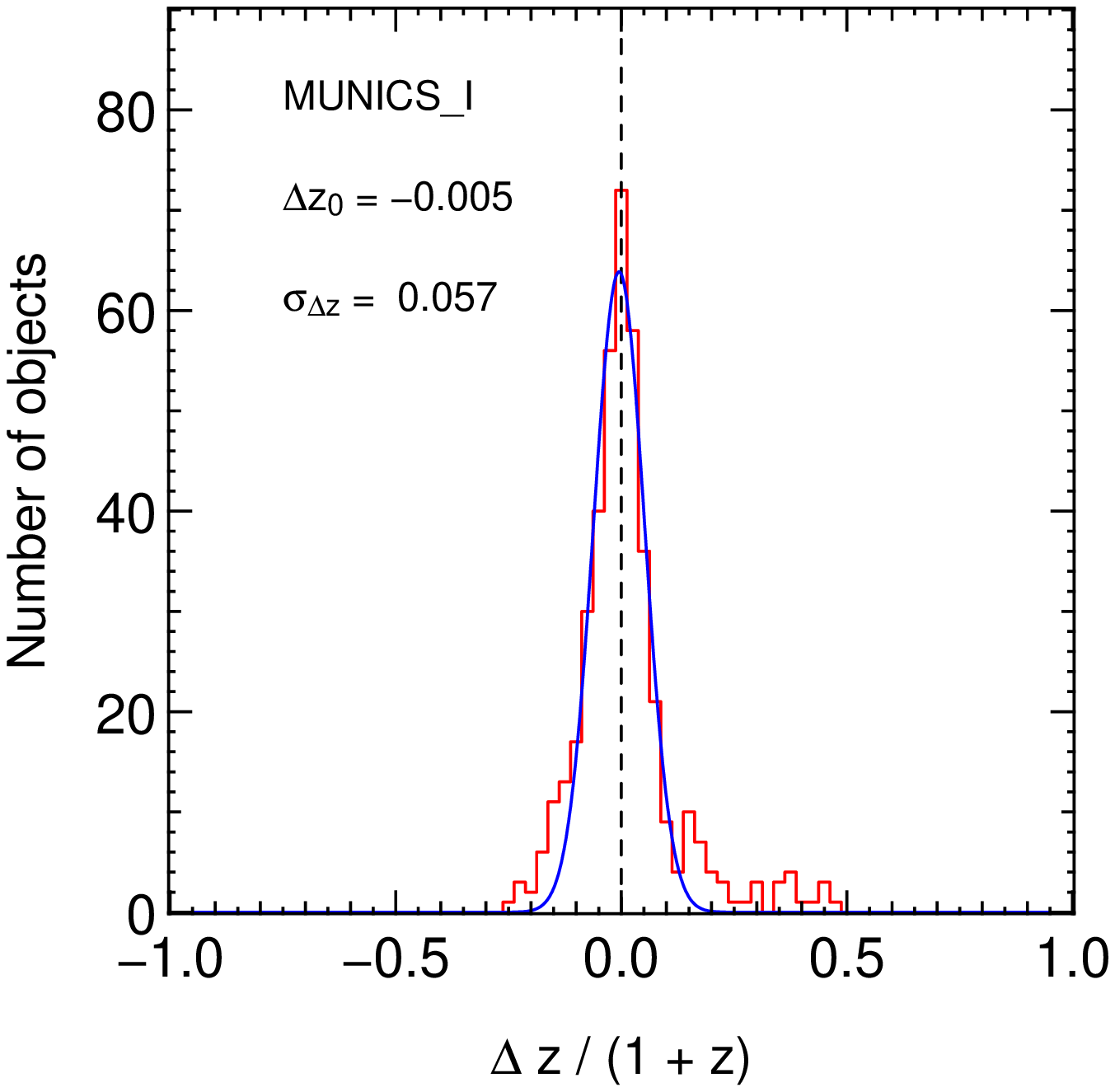,width=0.323\textwidth}
\hfill
\epsfig{figure=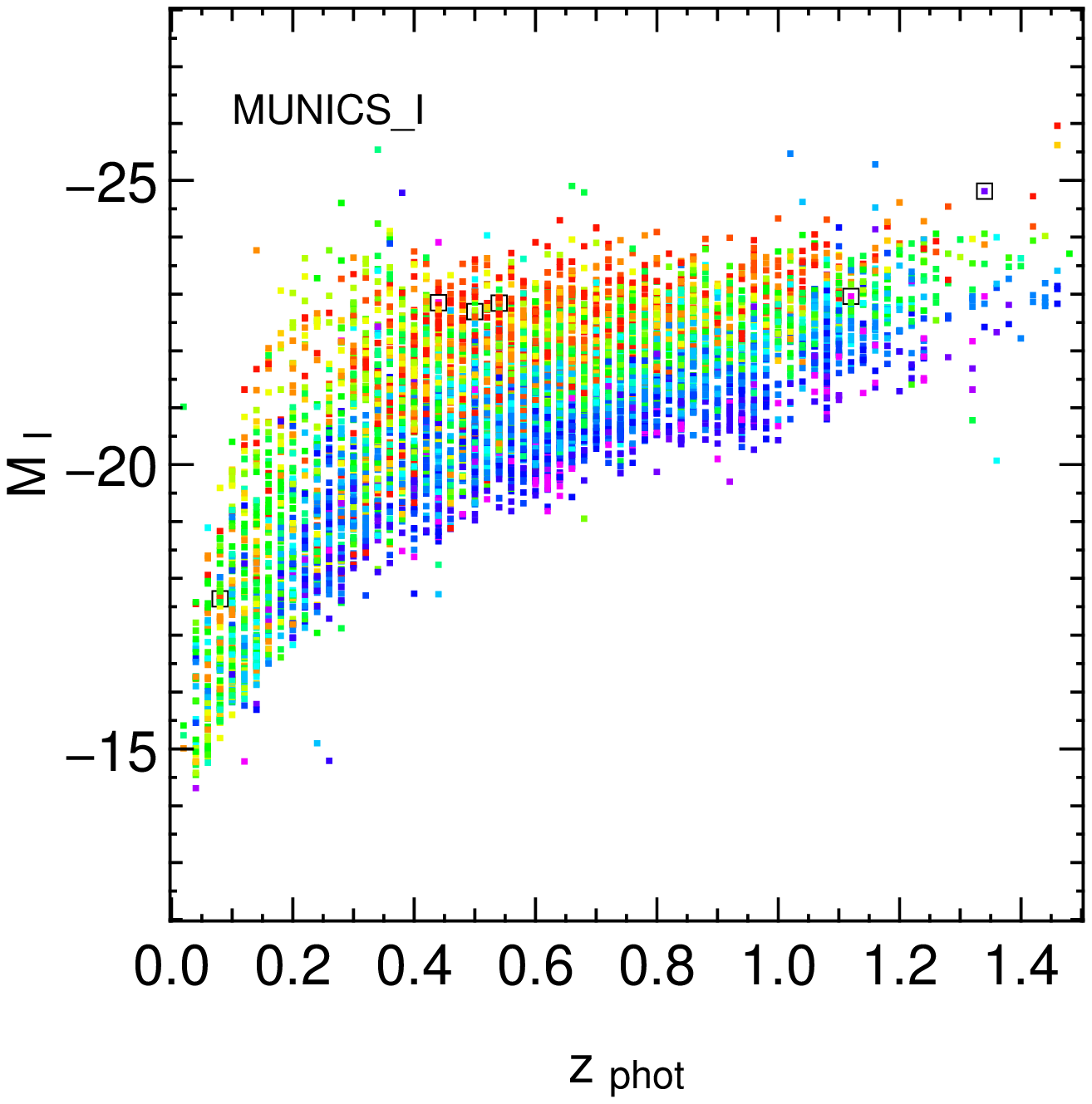,width=0.313\textwidth}
\caption{Photometric redshift for the $I$-selected catalogue. The
  left-hand panel shows a comparison of photometric and spectroscopic
  redshift for the different MUNICS fields (indicated by the different
  colours), while the middle panel gives the corresponding error
  histogram (red) and a Gaussian fit to it (blue). In the right-hand
  panel we present the distribution of absolute $I$ magnitudes $M_I$
  versus redshift $z_\mathrm{phot}$, where the different colours
  indicate different model SEDs ranging from early types (red) to late
  types (purple). Open symbols identify objects spectroscopically
  classified as AGN.}
\label{f:zphot}
\end{figure*}

During the last decade, observational research on galaxy formation and
evolution made a lot of progress. Especially two quantities and their
redshift evolution have gained considerable attention: The stellar
mass function of galaxies and the star formation rate (SFR). The
stellar mass function was measured out to redshifts of $z \sim 1.5$
(e.g.\ \citealt{Bell2003b, Dickinson2003, munics6, Fontana2004}),
while deep pencil beam surveys allowed to trace the SFR to even higher
redshifts (e.g.\ \citealt{CFRS96, Madau1996, Madau1998, Steidel1999,
Bouwens2004, fdfsfr}).

However, the total integrated SFR and its evolution with redshift does
not tell us about the contribution of star formation to the build up
of stellar mass for different galaxy masses. For example we cannot say
whether the general rise of the SFR to redshift one is produced by
high-mass or low-mass galaxies, and how much stellar mass galaxies of
different mass form during this period. \citet{Cowie1996} used
$K$-band luminosities and [OII] equivalent widths to investigate this
connection and noted an emerging population of massive, heavily star
forming galaxies at higher redshifts, a phenomenon they termed
`downsizing'.

A more direct measure of this connection is the `specific star
formation rate' (SSFR, \citealt{Guzman1997,BE00}) which is defined as
the SFR per unit stellar mass. This quantity allows us to explore the
relation between stellar mass and SFR directly. The SSFR has been
studied before, both locally \citep{PerezGonzalez2003,Brinchmann2004}
and at higher redshifts \citep{Guzman1997,BE00,Fontana2003,amanda}.
Our study, which relies on photometric redshifts, extends previous
work by investigating a large sample of galaxies at higher redshifts.

In this letter, we present measurements of the SSFR and its evolution
of redshift based on an $I$-band selected catalogue of more than 6000
field galaxies from the Munich Near-Infrared Cluster Survey (MUNICS),
allowing us to trace the change of the SSFR with cosmic time with high
statistical accuracy.

This letter is organised as follows. First we introduce the galaxy
sample in Section~\ref{s:sample} and describe our methods to derive
SFRs and stellar masses. In Section~\ref{s:ssfr} we present our
results on the SSFR, before we summarise our findings and discuss
their implications in Section~\ref{s:disc}.  Throughout this work we
assume $\Omega_m = 0.3$, $\Omega_\Lambda = 0.7$ and $H_0 = 70 \,
\mathrm{km} \, \mathrm{s}^{-1} \, \mathrm{Mpc}^{-1}$. All magnitudes
are given in the Vega system.

\section{The Galaxy Sample}
\label{s:sample}

Galaxies used in this study are drawn from the Munich Near-Infrared
Cluster Survey (MUNICS, \citealt{munics1}), a wide-area, medium deep
photometric and spectroscopic survey in the $BVRIJK$ bands covering an
area of about 0.3 square degrees down to $K \simeq 19$ and $R \simeq
24$ \citep{munics4}. In contrast to previous work on the $K$-selected
sample (``MUNICS\_K'', \citealt{munics3, munics2, munics6}), this work
is based on an $I$-band selected galaxy catalogue (``MUNICS\_I'')
which will be described in detail in a forthcoming paper (Feulner et
al. 2005, in preparation). Object detection and photometry was
performed using YODA \citep{YODA} in much the same way as for the
$K$-selected sample \citep{munics1}. We use the same sub-set of
high-quality fields as in \citet{munics3, munics2, munics6}. Stars are
excluded based on their spectral energy distributions (SEDs), leaving
6180 galaxies for further analysis.

Photometric redshifts are derived using the method described in
\citet{photred}. This is the same method also used on MUNICS\_K and
discussed in detail in \citet{munics2}. The photometric redshifts are
calibrated using the spectroscopic redshifts presented in
\citet{munics5}. Fig.~\ref{f:zphot} shows a comparison of photometric
and spectroscopic redshifts for MUNICS\_I as well as the distribution
of absolute $I$-band magnitudes $M_I$ versus redshift. The
distribution of redshift errors is similar to MUNICS\_K with a width
of $\Delta z/(1+z) = 0.057$.

We estimate the star formation rates (SFRs) of our galaxies from the
SEDs by deriving the luminosity at $\lambda = 2800 \pm 100$\AA\ and
converting it to an SFR as described in \citet{Madau1998} assuming a
Salpeter initial mass function (IMF; \citealt{Salpe55}).  We have
convinced ourselves that these photometrically derived SFRs are in
reasonable agreement with spectroscopic indicators for objects with
available spectroscopy. Note that since our bluest band is $B$, this
is an extrapolation for $z < 0.4$. Hence we restrict any further
analysis to redshifts $z > 0.4$, where the ultraviolet continuum at
$\lambda \simeq 2800$\AA\ is shifted into or beyond the $B$ band. The
SFR density as a function of redshift derived from our sample agrees
well with previous results and will be discussed in a future paper
(Feulner et al., 2005, in preparation).

Stellar masses are computed from the multi-colour photometry using a
method similar to the one used in \citet{munics6}. It is described in
detail and tested against spectroscopic and dynamical mass estimates
in Drory, Bender \& Hopp (\citeyear{masscal}). In brief, we derive
stellar masses by fitting a grid of stellar population synthesis
models by \citet{BC2003} with a range of star formation histories
(SFHs), ages, metalicities and dust attenuations to the broad-band
photometry. We describe star formation histories (SFHs) by a
two-component model consisting of a main component with a smooth SFH
$\propto \exp (-t/\tau)$ and a burst. We allow SFH timescales $\tau
\in [0.1,\infty]$~Gyr, metalicities $[\mathrm{Fe/H}] \in [-0.6,0.3]$,
ages between 0.5~Gyr and the age of the universe at the objects
redshift, and extinctions $A_V \in [0,1.5]$. The SFRs derived from
this model fitting is in good agreement with the ones from the UV
continuum. Note that we apply the extinction correction derived from
this fitting also to the SFRs.

\section{The Specific Star Formation Rate}
\label{s:ssfr}

\begin{figure*}
\epsfig{figure=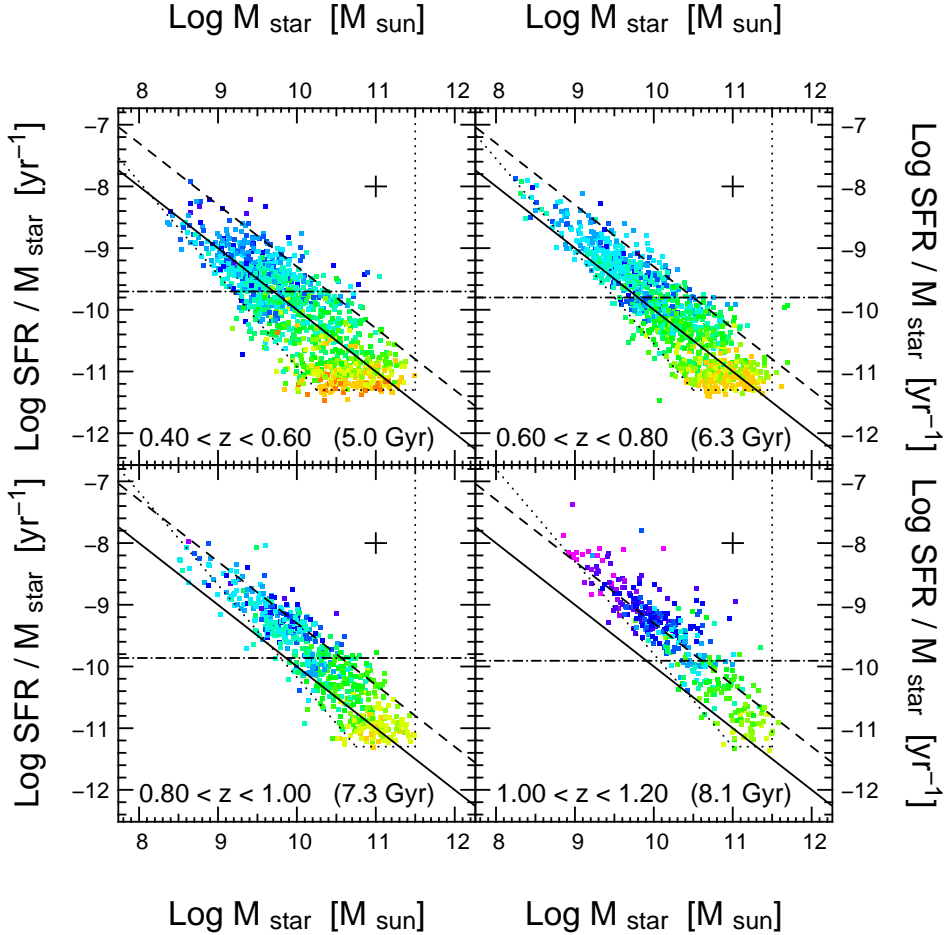,width=0.7\textwidth}
\caption{The SSFR as a function of stellar mass for MUNICS\_I. The
  solid and dashed lines correspond to SFRs of 1 $M_\odot \:
  \mathrm{yr}^{-1}$ and 5 $M_\odot \: \mathrm{yr}^{-1}$,
  respectively. While this is a good measure of the upper envelope of
  the majority of objects at $z \sim 0.5$, the point distribution
  shifts to higher SSFRs with increasing redshift. The dotted lines
  indicate the limits of the point distribution due to magnitude
  limits, the model SED set and the mass function (see the text for
  details). Objects are coloured according to the age of the CSP model
  fit to the photometry, ranging from 9~Gyr (red) to 0.05~Gyr
  (purple). The dot-dashed line is the SSFR required to double a
  galaxy's mass between each redshift epoch and today (assuming
  constant SFR); the corresponding look-back time is indicated in each
  panel. The error cross in each panel gives an idea of the typical
  errors.}
\label{f:ssfr}
\end{figure*}

We investigate the connection between SFR and stellar mass and its
evolution with redshift by considering the `specific star formation
rate' SSFR \citep{Guzman1997, BE00}, defined as the SFR per unit
stellar mass. In Fig.~\ref{f:ssfr} we show the SSFR as a function of
stellar mass for four different redshift bins from $z = 0.5$ to $z =
1.1$. The general shape is in very good agreement with a similar study
based on spectroscopic data \citep{amanda}.

Let us first understand the limits of the object distribution in this
diagram as indicated by the dotted lines. First, the sharp cut-off at
the high mass end at $\log M_\mathrm{star}/M_\odot \simeq 11.5$ is
produced by the high-mass cut-off of the stellar mass function (see
e.g.\ \citealt{munics6, Fontana2004}). Secondly, the lower limit at
$\log \mathrm{SSFR} \simeq -11.3$ is due to the fact that data points
fit by the same model SED occupy horizontal slices in the diagram,
with the reddest (oldest, least active) galaxies at the bottom and
subsequently bluer models along the distribution to higher values of
$\log \mathrm{SSFR}$. Finally, the limit of the point distribution to
the left of the diagram is due to a combination of the selection band
and the limiting magnitudes in all filters of the MUNICS survey. This
completeness limit will run parallel to lines of constant SFR in a
$B$-selected galaxy sample, but have a much steeper slope for
near-infrared selected galaxies.

\begin{figure}
\centerline{\epsfig{figure=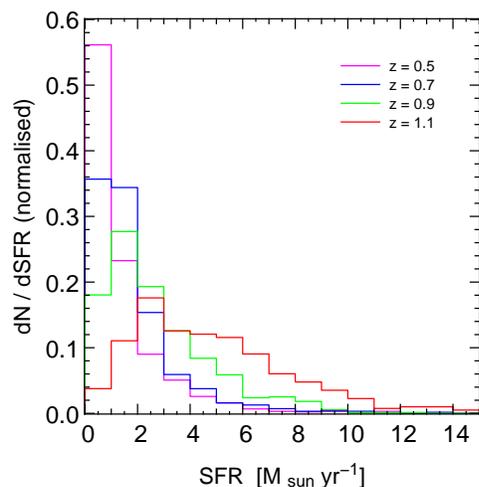,width=0.35\textwidth}}
\caption{Normalised histogram of the SFR for the four different
  redshift bins. The shift of the high-SFR cut-off to higher SFRs with
  increasing redshift is clearly visible. The individual histograms
  are divided by the number of objects in each redshift bin. Note that
  at higher redshifts incompleteness starts to cut away objects with
  low SFR as indicated in Fig.~\ref{f:ssfr}.}
\label{f:hist}
\end{figure}

The first result which can be derived from Fig.~\ref{f:ssfr} is that in our
optically-selected survey there is an upper bound on the SSFR (with a
few galaxies with very high SFRs which are likely starburst galaxies
or AGN). It runs parallel to lines of constant star-formation over a
wide range of masses $M \ga 10^9 M_\odot$ and at all redshifts,
meaning that this upper limit of the SFR does not depend on galaxy
mass. Furthermore, this maximum SFR is generally increasing with
increasing redshift for all stellar masses, from $\mathrm{SFR} \simeq
5 \: M_\odot \: \mathrm{yr}^{-1}$ at $z \simeq 0.5$ to $\mathrm{SFR}
\simeq 10 \: M_\odot \: \mathrm{yr}^{-1}$ at $z \simeq 1.1$. Note
that, while the lower part of the SSFR in the diagram is affected by
incompleteness, the constraints on the upper envelope are robust. This
is evident from Fig.~\ref{f:hist} where we show the histogram of the
SFR for the four different redshift bins, clearly showing the increase
of the maximum SFR with redshift.

We note that our sample might be missing highly obscured, heavily star
forming massive galaxies which could occupy the upper right part of
Fig.~\ref{f:ssfr}. Indeed, mid-infrared studies have shown that these
galaxies exist and that the upper bound of the SSFR is partly a
selection effect due to their dust content \citep{Hammer2001,
Franceschini2003}. However, we can conclude from their number density
that they contribute at most 10\% to the field galaxy population (the
objects studied by \citet{Franceschini2003} have 25\% of the number
density of our sample, but their optical data go roughly 2~mag deeper
than ours). Thus, even if our sample should miss these galaxies, our
conclusions still hold for the larger part of the field galaxy
population. While differences in extinction between galaxies of
different mass might influence the shape of the upper bound, its
existence and observed change are robust to these differential
effects.

Hints for a shift of this upper envelope to higher SSFRs with redshift
were already noted by \citet{BE00} and \citet{amanda} from smaller
galaxy samples, but our large sample of more than 6000 galaxies
allows to constrain this change in a much more robust way.

Furthermore, we can study the distribution of the ages of the model
stellar populations in Fig.~\ref{f:ssfr}. It is clear that the most
massive galaxies contain the oldest stellar populations at all
redshifts, with ages close to the age of the universe at each epoch.

Finally we indicate the SSFR needed to double a galaxy's stellar mass
between the epoch of observation and today (assuming a constant
SFR). Clearly, the most massive galaxies are well below this line at
all redshifts, indicating that they formed the bulk of their stars at
earlier times, in agreement with the age distribution discussed
above. This also means that star formation contributes much more to
the mass build-up of less massive galaxies than to high-mass
systems. While between redshifts $z=1$ and $z=0$ the mass of a
$10^{11} M_\odot$ system would typically change by $\sim 40\%$ due to
star formation, the mass of $10^{10} M_\odot$ galaxies would grow by a
factor of $\sim 5$ and that of $10^9 M_\odot$ systems by a factor of
$\sim 40$. This example assumes a constant SFR of $\dot{\varrho}_\star
= 5 \: M_\odot \: \mathrm{yr}^{-1}$ over a period of 7.7~Gyr which, as
will be shown below, is likely to be unrealistic (at least for the
lower-mass systems).

\section{Summary and Conclusions}
\label{s:disc}

We have presented the specific star formation rate (SSFR) as a
function of stellar mass and redshift for a large sample of more than
6000 $I$-band selected galaxies. The SSFR decreases with mass at all
redshifts, although we might not detect highly obscured galaxies. The
low values of the SSFR of the most massive galaxies suggests that most
of these massive systems formed the bulk of their stars at earlier
epochs. Furthermore, stellar population synthesis models show that
these most massive systems contain the oldest stellar populations at
all redshifts. This is in agreement with the detection of old, massive
galaxies at redshifts $1 \la z \la 2$ \citep{tesis1, Cimatti2004}.

In our optically-selected sample, there is an upper bound to the SSFR
of the majority of field galaxies which is parallel to lines of
constant star formation rate (SFR). This upper limit on the SFR is
independent of stellar mass, but increases with redshift from
$\mathrm{SFR} \simeq 5 \: M_\odot \: \mathrm{yr}^{-1}$ at $z \simeq
0.5$ to $\mathrm{SFR} \simeq 10 \: M_\odot \: \mathrm{yr}^{-1}$ at $z
\simeq 1.1$.

We can also infer from Fig.~\ref{f:ssfr} that star formation in lower mass
galaxies cannot proceed at constant SFR for a long time: All galaxies
above the dot-dashed line in the diagram have the potential to double
their stellar mass between the epoch of observation and today
(assuming a constant SFR). While lower mass galaxies at low redshift
tend to be gas rich, there is a large spread in measured
gas-to-stellar-mass fractions \citep{Mateo1998, PerezGonzalez2003,
Kannappan2004}. However, very gas-rich systems are rare
\citep{Davies2001}, i.e.\ the majority of these galaxies does not have
huge gas supplies, which might lead us to believe that low-mass
galaxies cannot exhibit constant star formation over longer
time-scales, but show variable star formation histories, like the ones
derived for the -- even lower mass -- dwarf galaxies in the Local
Group (see e.g.\ \citealt{Mateo1998, Tosi2001, Grebel2004} for
reviews). Due to the degeneracy of different star formation histories
in colour space, it is not possible to say from our data whether we
see these galaxies in the process of formation or during one of
multiple episodes of active star formation. However, it is likely that
we pick them up during an active phase of star formation. Also, it is
clear from the completeness limits that we cannot detect low-mass
galaxies with low SSFR.

Considering the high-mass end, we can try to draw some conclusions
about the contributions of star formation and merging to the change of
stellar mass. Between redshifts $z \simeq 1.1$ and $z \simeq 0.5$, the
characteristic mass of the cut-off of the galaxies' stellar mass
function changes by $\Delta \log M \simeq 0.15$~dex \citep{munics6,
Fontana2004, Conselice2004}. For a $M_\star = 10^{11} \: M_\odot$
stellar mass galaxy, a constant SFR of $\dot{\varrho}_\star = 5 \:
M_\odot \: \mathrm{yr}^{-1}$ over a period of time of $\Delta t =
3.1$~Gyr (the difference in time between these redshift values),
yields a growth in stellar mass of $\Delta M_\star \simeq 2 \cdot
10^{10} \: M_\odot$, or $\Delta \log M_\star \simeq
0.1$~dex. Considering the uncertainty of the results and our lack of
knowledge about the star formation histories of these galaxies, we
cannot really decide about the relative importance of star formation
and merging. We note, however, that our results on the growth of
stellar mass in massive galaxies does not require a substantial
contribution of merging over the redshift range $0 \; \la \; z \; \la
\; 1$.

Overall it is clear that there is a marked difference between the star
formation histories of low-mass and high-mass galaxies in agreement
with findings from the stellar populations of today's galaxies
\citep{Heavens2004, Thomas2004}.

\section*{Acknowledgments}

We thank the anonymous referee for his comments which helped to
improve the presentation of this letter. The authors would like to
thank the staff at Calar Alto Observatory for their
support. Furthermore, we thank Amanda Bauer for discussion and Jan
Snigula for help with the colour figures. We acknowledge funding by
the DFG (SFB 375). This research has made use of NASA's Astrophysics
Data System (ADS) Abstract Service.

\bibliographystyle{mn2e}
\bibliography{mnrasmnemonic,literature}

\label{lastpage}

\end{document}